\documentclass[11pt]{article}
\usepackage{amsfonts}
\usepackage{graphicx}
\usepackage{amsmath}
\usepackage{amssymb}

\begin{document}

\title{Selection rules for light-by-light scattering in strong magnetic field%
}
\author{A.E.Shabad \\
shabad@lpi.ru, P. N. Lebedev Physical Institute, 53 Leninskiy prospekt,\\
119991, Moscow, Russia\\
Tomsk State University, 36 Lenin Prospekt, 634050, Tomsk, Russia}
\maketitle

\begin{abstract}
Selection rules that follow from CP- and 4-momentum conservation are listed
for head-on light-by-light scattering in strong magnetic field taking into
account nontrivial dispersion laws of different photon eigenmodes.
\end{abstract}

\section{Introduction}

Recently R. Baier, A. Rebhan, and M. W\"{o}dlinger \cite{Baier} calculated
cross sections of light-by-light scattering as functions of an external
magnetic field $\mathbf{B}$\ for low-energy long-wave photons, basing on the
Heisenberg-Euler Lagrangian -- considered in QED, in scalar electrodynamics
and in the theory with charged vector bosons. Calculations are done for
special cases of equal-wave-length photons colliding head-on, when the
external field is either parallel or perpendicular to the incoming photon
direction, so that the three-momenta of the two colliding photons are
subject to the relation $\mathbf{k}_{1}=-\mathbf{k}_{2}.$This kinematical
restriction cannot, generally, be avoided by a Lorentz transformation to an
arbitrary frame, since -- when $\mathbf{B\nshortparallel k}_{1,2}$ -- there
is no invariance under the boost in the direction of the photon propagation.
Nevertheless, the head-on collision case corresponds to the realistic
experimental situation, where two laser beams collide, so that in the
laboratory frame their center-of-mass is at rest.

In the present note we stress that the kinematics of the reaction cannot be
borrowed without change from the vacuum case, because the presence of the
magnetic field does affect it. Relation between the energy and momentum of
the photon is not just $\omega =k,$ but the energy of each photon may depend
in two ways on the angle, which the direction of its momentum makes with the
magnetic field, $\omega _{\text{ordinary,extraordinary}}(k^{2},\left(
\mathbf{k\cdot B}\right) ^{2}),$ reflecting the anisotropy of the effective
"optical medium" and birefringence$.$ (Within the local approximation kept
to in \cite{Baier}, and here as well, the dependence on the angle is very
simple, see Eqs. (13), (25), (26) in \cite{Usov2011}).

The photons that are inside the magnetic field are classified according to
the eigenmodes. (If one is willing to consider scattering of photons falling
from outside of the region occupied by the magnetic field, one should also
take into account their reflection and refraction at the border.)\textbf{\ }%
Definite laws of propagation -- the corresponding refraction indices,
propagation speeds \textit{etc.}-- are associated with photon eigenmodes,
which are not the photons just transversely polarized. (The polarizations of
eigenmodes are established in \cite{batalin} and partially in Ref. \cite%
{adler}). For the general angle between $\mathbf{k}$ and $\mathbf{B}$ the
so-called ordinary wave is transverse (this is mode-3 in classification of (%
\cite{batalin}), its electric field is orthogonal to the plane spanned by $%
\mathbf{k}$ and $\mathbf{B}$, while the extraordinary wave (mode-2) is not
(its electric vector belongs to that plane). So, the scattered photons are
not all transverse even when the incoming photons are parallel or orthogonal
to the magnetic field.

In the next Section I illustrate the influence of the anisotropic dispersion
laws by considering relations among momenta and scattering angle of the
photons within the same configurations of the incident photons and the
magnetic fields as the ones considered in \cite{Baier}. I did not find
kinematical bans, analogous to those known for photon splitting, but the
bans due to CP-conservation like in Adler's work \cite{adler} are expected.
Essential also is the change of the photon wave-length depending (for
parallel incidence) or not depending (for perpendicular incidence) on the
scattering angle.

\subsection{Selection rules}

The energy-momentum conservation reads

\begin{eqnarray}
\mathbf{k}_{1}+\mathbf{k}_{2} &=&\mathbf{k}_{3}+\mathbf{k}_{4}  \label{1} \\
\omega _{1}+\omega _{2} &=&\omega _{3}+\omega _{4}.  \label{2}
\end{eqnarray}%
Energy of each photon $i=1,2,3,4$ may belong to one of the modes $a=2$,$3$
(sometimes also mode-1, $a=1,$ comes into play, see below), and it depends
on its momentum and on orientation of the latter with respect to the
magnetic field:%
\begin{equation}
\omega _{i}=\omega ^{(a_{i})}(k_{i}^{2},(\mathbf{k}_{i}\mathbf{\cdot B}%
)^{2}),\quad k_{i}=|\mathbf{k}_{i}|.  \label{3}
\end{equation}%
We say that we face the "center-of-mass" configuration if additionally the
relation%
\begin{equation}
\mathbf{k}_{1}+\mathbf{k}_{2}=\mathbf{k}_{3}+\mathbf{k}_{4}=0  \label{4}
\end{equation}%
is obeyed (photons of equal wave-length are colliding head-on). In this
configuration $\mathbf{k}_{4}$ lies in the plane spanned by the vectors $%
\mathbf{k}_{1}=-\mathbf{k}_{2}$ and $\mathbf{k}_{3},$ because $\mathbf{k}%
_{4}=-\mathbf{k}_{3}.$ In other words, the initial and final reaction planes
coincide. (I don't know if the same statement is true for the general
configuration under conditions of the lack of Lorentz and rotational
invariance that prevents one from passing to the general configuration by
changing the reference frame.) I shall confine myself to the "center-of-mass
configuration" (\ref{4}) in what follows.

With the account of (\ref{4}) Eq. (\ref{3}) may be written as
\begin{equation}
\omega _{i}=\omega ^{(a_{i})}(k_{i}^{2},(\mathbf{k}_{i}\mathbf{\cdot B}%
)^{2}),\quad k_{1,3}=k_{2,4},\quad (\mathbf{k}_{1,3}\mathbf{\cdot B})^{2}=(%
\mathbf{k}_{2,4}\mathbf{\cdot B})^{2},  \label{4'}
\end{equation}%
i.e. $\mathbf{k}_{1}$ represents the incident photons, and $\mathbf{k}_{3}$
represents the scattered ones. (When the two incoming photons belong to the
same mode, $a_{1}=a_{2},$ the configuration considered may be described as
scattering of equal-energy photons, because then $\omega ^{(a_{1})}(k_{1},(%
\mathbf{k}_{1}\mathbf{\cdot B})^{2})=\omega ^{(a_{2})}(k_{2},(\mathbf{k}_{2}%
\mathbf{\cdot B})^{2})$ thanks to the relation $\mathbf{k}_{1}=-\mathbf{k}%
_{2}$. I shall not restrict myself to this case in what follows, however).

The dispersion laws for mode-2 and mode-3 waves in the original
classification of Refs. (\cite{batalin}), $a=2,3$ are in the long-wave,
low-frequency approximation governed by the Heisenberg-Euler Lagrangian
(HEL) \cite{Usov2011}
\begin{equation}
\omega ^{(2,3)}=\left( \left( \frac{(\mathbf{k\cdot B})}{B}\right)
^{2}+\left( k^{2}-\left( \frac{(\mathbf{k\cdot B})}{B}\right) ^{2}\right)
c^{(2,3)}\right) ^{1/2},  \label{displaws}
\end{equation}%
where the coefficients $c^{(2,3)}$\ are known dimensionless functions of the
external field expressed in terms of first- and second-order derivatives of
HEL. It agrees with the causality that $c^{(2,3)}<1.$ While investigating
the selection rules it may be important that $c^{(3)}>c^{(2)}.$

The energy conservation relations (\ref{2})
\begin{equation}
\omega ^{(a_{1})}(k_{1},(\mathbf{k}_{1}\mathbf{\cdot B})^{2})+\omega
^{(a_{2})}(k_{1},(\mathbf{k}_{1}\mathbf{\cdot B})^{2})=\omega
^{(a_{3})}(k_{3},(\mathbf{k}_{3}\mathbf{\cdot B})^{2})+\omega
^{(a_{4})}(k_{3},(\mathbf{k}_{3}\mathbf{\cdot B})^{2})  \label{selection}
\end{equation}%
are fraught with dynamic selection rules that may forbid many of the sixteen
transitions (four initial by four final polarization states). Besides, there
is the parity ban (cf. \cite{adler}) for the transitions with the
participation of a total odd number of mode-3 photons in initial and final
states, since the mode-3 vector-potential is a pseudovector. Hence mode-2
photon may appear only even number of times among the four photons
participating in the reaction. The CP-selection rules derived thereof from
this general consideration manifest themselves by calculations of \ Ref.\cite%
{Baier}

\subsubsection{Perpendicular incidence}

Consider first the simplest case when the incoming photon momenta are
perpendicular to the magnetic field, $\mathbf{k}_{1,2}\perp \mathbf{B.}$ So
are the outgoing momenta since they lie in the same (reaction) plane. In
this case the longitudinally polarized component of the electric field in
the extraordinary mode-2-wave disappears. The modes 2 and 3 are mutually
orthogonal, both transverse, electromagnetic waves. Their dispersion laws
are
\begin{equation}
\omega _{1,2}=\omega ^{(a_{1,2})}(k_{1},0),\quad \omega _{3,4}=\omega
^{(a_{3,4})}(k_{3},0).  \label{k''}
\end{equation}%
The energy-conservation relations (\ref{selection}) take the form%
\begin{equation}
\omega ^{(a_{1})}(k_{1},0)+\omega ^{(a_{2})}(k_{1},0)=\omega
^{(a_{3})}(k_{3},0)+\omega ^{(a_{4})}(k_{3},0).  \label{selperp}
\end{equation}%
To see what rules these equations imply let me consider first one out of 16
transitions, when two photons of mode-2 collide to produce two photons of
mode-3: ($2,2)\rightarrow (3,3)$ . Eq. (\ref{selperp}) requires
\begin{equation*}
\omega ^{(2)}(k_{1},0)=\omega ^{(3)}(k_{3},0),
\end{equation*}%
or using (\ref{displaws})%
\begin{equation}
k_{1}\left( c^{(2)}\right) ^{1/2}=k_{3}\left( c^{(3)}\right) ^{1/2}.
\label{k1>k3}
\end{equation}%
This relation establishes an obligatory connection between the wave lengths
of the incoming and outgoing photons. Since $c^{(3)}>c^{(2)},$ the outgoing
wave has a longer length, $k_{3}<k_{1}.$ For the opposite process $%
(3,3)\rightarrow (2,2)$ the selection rule
\begin{equation}
k_{1}\left( c^{(3)}\right) ^{1/2}=k_{3}\left( c^{(2)}\right) ^{1/2}
\label{k3>k1}
\end{equation}%
implies the opposite inequality: $k_{3}>k_{1}.$

On the contrary, the transitions when two different-mode photons turn into
also two different-mode photons, $(3,2)\rightarrow (3,2),$ demand that $%
k_{1}=k_{3}$, because Eq. (\ref{selperp}) becomes in this case

\begin{equation*}
k_{1}\left( c^{(2)}\right) ^{1/2}+k_{1}\left( c^{(3)}\right)
^{1/2}=k_{3}\left( c^{(2)}\right) ^{1/2}+k_{3}\left( c^{(3)}\right) ^{1/2}.
\end{equation*}

It remains to consider transitions when all the four photons are of the same
polarization, $(2,2)\rightarrow (2,2)$ and $(3,3)\rightarrow (3,3)$%
\begin{equation}
\omega ^{(2,3)}(k_{1},0)+\omega ^{(2,3)}(k_{1},0)=\omega
^{(2,3)}(k_{3},0)+\omega ^{(2,3)}(k_{3},0).
\end{equation}%
\begin{equation*}
k_{1}\left( c^{(2,3)}\right) ^{1/2}+k_{1}\left( c^{(2,3)}\right)
^{1/2}=k_{3}\left( c^{(2,3)}\right) ^{1/2}+k_{3}\left( c^{(2,3)}\right)
^{1/2}.
\end{equation*}%
Therefore such process requires, as before, that $k_{1}=k_{3}.~$

Other transitions $(2,2)\leftrightarrow (2,3),$ \ $(3,3)\leftrightarrow
(2,3) $ are forbidden since they would violate parity.

To conclude this Subsection we state that the light-by-light scattering in
the "center-of-mass" configuration across the magnetic field requires that
the wave-length should conserve, $k_{1}=k_{3},$ when none or one photon
changes its polarization, but it should not, $k_{1}\gtrless k_{3},$ when two
photons both change their polarization. Processes, where the photon, whose
polarization is given by mode-3, is involved once or thrice are
parity-impossible.

\subsubsection{Parallel incidence}

When two initial photons are oriented parallel to $\mathbf{B,}$ (and the
final are not) the energy conservation (\ref{selection}) gives%
\begin{equation}
\omega ^{(a_{1})}(k_{1},k_{1}^{2}B^{2})+\omega
^{(a_{2})}(k_{1},k_{1}^{2}B^{2})=\omega ^{(a_{3})}(k_{3},(\mathbf{k}_{3}%
\mathbf{\cdot B})^{2})+\omega ^{(a_{4})}(k_{3},(\mathbf{k}_{3}\mathbf{\cdot B%
})^{2}).  \label{selpar}
\end{equation}%
The falling photons may belong either to (transverse in this case) mode-1 or
to ever transverse mode-3, since mode-2 in disappears for parallel
propagation (see second reference in \cite{batalin}). Let all the four
involved photons belong to mode-3: $a_{1,2,3,4}=3,$ that is we consider the
process $(3,3)\rightarrow (3,3).$ Then (\ref{selpar}) becomes%
\begin{eqnarray}
\omega ^{(3)}(k_{1},k_{1}^{2}B^{2}) &=&\omega ^{(3)}(k_{3},(\mathbf{k}_{3}%
\mathbf{\cdot B})^{2}))\quad or  \notag \\
\omega ^{(3)}(k_{1},k_{1}^{2}B^{2}) &=&\omega ^{(3)}(k_{3},k_{3}^{2}(B\cos
\theta )^{2})).  \label{selpar2}
\end{eqnarray}%
Here $\theta $ is the scattering angle of Eq. (14) in \cite{Baier}. With the
help of (\ref{displaws}) we obtain for $\left( \ref{selpar2}\right) $%
\begin{eqnarray}
k_{1} &=&\left( k_{3}^{2}(\cos \theta )^{2}+\left( k_{3}^{2}-k_{3}^{2}(\cos
\theta )^{2}\right) c^{(3)}\right) ^{1/2},  \notag \\
k_{1} &=&k_{3}\left( \cos ^{2}\theta +c^{(3)}\sin ^{2}\theta \right) .
\label{2par}
\end{eqnarray}%
This is a definite kinematical relation between the scattering angle and the
ratio of the initial and final wave-lengths parameterized by the magnetic
field hidden in $c^{(3)}$, -- obligatory for the chosen process to be
permitted. Here the present case of parallel incidence differs from the
perpendicular incidence of the previous item, where this ratio was fixed,
but the scattering angle remained unrestricted. It follows from (\ref{2par})
and the causality $c^{(3)}<1$ that $k_{1}<k_{3},$ the outgoing waves are
shorter than the incoming ones$.$

Let now two incident photons belong both to mode-3, $a_{1,2}=3,$ while the
scattered photons to mode-2: $a_{3,4}=2.$ Then for the process $%
(3,3)\rightarrow (2,2)$ (\ref{selpar}) becomes%
\begin{equation*}
\omega ^{(3)}(k_{1},k_{1}^{2}B^{2})=\omega ^{(2)}(k_{3},k_{3}^{2}(B\cos
\theta )^{2})).
\end{equation*}%
\begin{equation}
k_{1}=k_{3}\left( \cos ^{2}\theta +c^{(2)}\sin ^{2}\theta \right)
,k_{1}<k_{3}.  \label{3par}
\end{equation}

This relation is of the same type as (\ref{2par}).

More special are the cases where one or both incident photons belong to
mode-1. These are $(1,3)\rightarrow (2,3),$ $(1,1)\rightarrow (3,3),$ $%
(1,1)\rightarrow (2,2).$ Let us take the process $(1,3)\rightarrow (2,3)$ to
begin with. The dispersion law for mode-1 under parallel propagation is just
the vacuum dispersion law $\omega ^{(1)}(k,k^{2}B^{2})=k.$ The energy
conservation reads
\begin{equation*}
\omega ^{(1)}(k_{1},k_{1}^{2}B^{2})+\omega
^{(3)}(k_{1},k_{1}^{2}B^{2})=\omega ^{(3)}(k_{3},k_{3}^{2}(B\cos \theta
)^{2}))+\omega ^{(2)}(k_{3},k_{3}^{2}(B\cos \theta )^{2})).
\end{equation*}%
\begin{equation}
k_{1}\left( 1+\cos ^{2}\theta +c^{(3)}\sin ^{2}\theta \right) =k_{3}\left(
2\cos ^{2}\theta +\left( c^{(3)}+c^{(2)}\right) \sin ^{2}\theta \right)
,\quad k_{1}<k_{3}.
\end{equation}

Analogously, the process $(1,1)\rightarrow (3,3)$ requires that
\begin{equation}
k_{1}=k_{3}\left( \cos ^{2}\theta +c^{(3)}\sin ^{2}\theta \right) ,\quad
k_{1}<k_{3}
\end{equation}%
and the process $(1,1)\rightarrow (2,2)$ that%
\begin{equation}
k_{1}=k_{3}\left( \cos ^{2}\theta +c^{(2)}\sin ^{2}\theta \right) ,\quad
k_{1}<k_{3}
\end{equation}

.

\subsection{Conclusion}

We have established selections rules for the head-on photon-photon
collisions in the vacuum filled by a strong magnetic field, which is
parallel or perpendicular to the axis, on which the momenta of the two
incoming photons lie. Some combinations of the initial and final photon
polarization eigenstates proved out to be excluded by the parity
conservation. All the rest are kinematically permitted, but the wavelengths
of the final photons for certain combinations of initial and final
polarizations differ from those of initial photons. This difference may
become significant for the magnetic fields of the critical order $\ B_{cr}=%
\frac{m^{2}}{e}=4.4\cdot 10^{13}$G. This means that the device registering
the photons resulting from two laser beams collision should be tuned to an
energy different from the energy of the colliding photons. This prescription
results from taking into account of different dispersion laws in different
photon modes. The quantitative part of the consideration fits low-frequency,
long-wave photons because it is based on the local approximation as given by
HEL.

\section*{Acknowledgements}

Supported by RFBR under Project 17-02-00317, and by the TSU Competitiveness
Improvement Program, by a grant from \textquotedblleft The Tomsk State
University D. I. Mendeleev Foundation Program\textquotedblright .

\bigskip

\bigskip

\end{document}